# Modeling of ASD/TD Children's Behaviors in Interaction with a Virtual Social Robot During a Music Education Program Using Deep Neural Networks

Armin Tandiseh[1], Morteza Memari[1], Alireza Taheri[1*]

[1]Social and Cognitive Robotics Laboratory, Sharif University of Technology, Tehran, Iran

* Corresponding author: artaheri@sharif.edu , Tel: +982166165531

**Abstract**

Autism Spectrum Disorder (ASD) is a neurodevelopmental condition characterized by difficulties in communication and social interaction, along with repetitive and restricted behaviors. Effective interventions are crucial to improving skills and the quality of life for individuals with ASD. Music is a promising educational/therapeutic approach that can help enhance social communication, emotional regulation, and sensory integration in children with ASD. Combining music-based games with virtual reality (VR) technology offers an immersive, multisensory experience that allows children with ASD to explore and interact with their environment in a safe and controlled way. This research aimed to develop an intelligent system to evaluate performance and extract behavioral models for children with ASD and neurotypical (TD) children by interacting with a virtual social robot in a music education program using deep neural networks. The system has two main features: 1) it distinguishes between neurotypical children and those with ASD based on their behavior, and 2) generates behaviors resembling those of neurotypical or ASD children in similar situations using deep learning. Intelligent systems that identify complex patterns and simulate behavior can aid in diagnosis, therapist training, and understanding the disorder. Using data from a previous study at the Social and Cognitive Robotics Laboratory of Sharif University of Technology (including the usable data of 9 ASD and 21 TD participants), the system achieved an accuracy of 81% and sensitivity of 96% in distinguishing neurotypical children from those with ASD using both impact data and motion signals. A transformer-based network was designed to reproduce children's behaviors. Experts in the field struggled to differentiate real behaviors from reproduced ones, with an accuracy of 53.5% and agreement of 68%, indicating the model's success in simulating realistic behaviors.

## I. Introduction

Autism Spectrum Disorder (ASD) is a complex neurodevelopmental condition that typically emerges within the first three years of a child's life. It is primarily characterized by persistent challenges in social communication and interaction, alongside restricted, repetitive patterns of behavior and interests. While symptoms are often apparent in early childhood, they can be more subtle and may be diagnosed later in children with milder forms of the disorder. Early diagnosis is critical, as timely intervention and support can profoundly improve a child's skills and overall quality of life [1]. The global prevalence of ASD is increasing, with recent studies from the Centers for Disease Control and Prevention showing a rise from 1 in 44 children in 2018 [2] to 1 in 36 in 2020 [3], underscoring the growing need for effective screening and intervention tools.

Intelligent modeling systems represent a promising new frontier in addressing this need [4, 5]. These systems can identify complex behavioral patterns that are often difficult for human observers to detect, thereby simplifying the diagnostic process for clinicians [6, 7]. Furthermore, they offer a unique capability to simulate the behaviors of individuals with ASD, providing a safe and effective training environment for new therapists without the necessity of a real child's presence. This technology not only aids in refining diagnostic accuracy but also enhances the educational and training opportunities for professionals, allowing them to gain a deeper understanding of the diverse facets of ASD [8, 9].

The primary objective of this research is to develop an intelligent system that can evaluate and model the behaviors of both typically developing (TD) children and those with ASD. This system operates within a virtual reality-based music education program, using advanced deep neural networks to extract and compare behavioral patterns. Our research aims to achieve two main goals: 1) to accurately differentiate between typically developing children and those with ASD based on their performance and behavior; and 2) to generate realistic, child-like behaviors for both groups, effectively creating a virtual representation for training and research purposes.

## II. Related Works

Research into Autism Spectrum Disorder with the help of technological tools has largely been categorized into three main areas: screening, treatment, and modeling. Each of these fields has seen significant advancements, particularly with the integration of artificial intelligence (AI) [10, 11].

### 1. Screening

Early screening for ASD has traditionally relied on clinical methods that involve behavioral observations and structured interviews. While these methods are considered the gold standard due to their personalized and insightful nature, they can be challenging due to potential human error and the need for highly specialized expertise. The increasing use of AI offers a powerful complementary tool, showing a significant shift in research towards advanced technologies to enhance or bypass the limitations of traditional screening [12].

AI-based models, especially deep learning, have demonstrated promising results by analyzing complex datasets. For instance, Li et al. used functional magnetic resonance imaging (fMRI) and deep learning to identify brain markers of ASD with 87.1% accuracy [13]. Other studies have utilized AI for behavioral analysis from naturalistic observations. Ghafghazi et al. developed an AI platform to automatically collect and analyze data, supporting personalized treatment plans and early detection of developmental disorders [14]. Carette et al. used an LSTM network to analyze eye-tracking data, achieving 83% accuracy in diagnosing ASD with a 95% confidence level [15]. Similarly, Tao et al. introduced the SP-ASDNet, a CNN-LSTM hybrid model, which classified individuals with ASD based on scan paths from images with 74.22% accuracy [16].

Several studies have explored different data modalities. Ramesh et al. used machine learning to analyze speech data, finding that Logistic Regression and Random Forest models reached 75% accuracy [17]. Farooq et al. applied federated learning to detect ASD in children and adults, reporting accuracies of 98% and 81% respectively, highlighting AI's potential to address data privacy concerns while maintaining high diagnostic accuracy [18]. Reddy et al. used deep learning to analyze facial images, with their best model achieving 88.33% accuracy [19]. Finally, a study by Serna Aguilera et al. used a CNN and Transformer hybrid model on video data of children's responses to sensory stimuli, achieving 81.48% accuracy [20].

2. Treatment

In the realm of treatment, AI and robotics are being used to enhance the effectiveness of therapeutic interventions. Studies have shown that interactive games and robots can significantly accelerate the treatment process [21, 22]. Scassellati et al. investigated the use of an autonomous social robot to improve social skills in children with ASD, reporting significant improvements in social interactions and a reduced need for guidance after a month-long intervention [23]. Giannetti et al. also used a humanoid robot to improve joint attention skills, noting that structured stimuli from the robot led to significant improvements and increased children's engagement and motivation [24].

Music therapy/education, in particular, has been identified as an effective intervention for ASD [25, 26]. It can improve social communication [27], emotional regulation [28], and motor skills [29]. Both non-personalized and personalized music therapy/education approaches exist, with personalized interventions showing greater effectiveness by adapting to the child's specific needs and reactions [30].

3. Modeling

The final category, modeling, focuses on creating behavioral simulations. This field is particularly relevant for training therapists without the need for real-child interaction. Baraka et al. designed a model based on the ADOS-2 standard to simulate the behaviors of children with ASD on a robot [8]. This model could generate random behaviors based on a child's age, speech ability, and ASD severity. Other studies have focused on modeling human behavior in complex environments. Bobu et al. introduced the Less method as an alternative to the traditional Boltzmann model to better predict human-like paths [31]. Pearce et al. used a combination of Behavioral Cloning and Reinforcement Learning to model human behavior in a video game, demonstrating the ability to predict and replicate realistic human actions [9].

Our research aligns with these ongoing efforts by addressing a significant gap in the literature. To the best of our knowledge, no prior study has focused on modeling and comparing the behaviors of both typically developing children and those with ASD during a virtual reality-based music education program. By integrating hit data and motion signals in a xylophone/drum-playing VR-based educational study, we aim to provide an automatic framework for ASD screening and behavioral modeling, paving the way for more effective clinical interventions and therapist training.

## III. Methodology

### 1. Game Environment and Protocol

This study involved 14 children with ASD (mean age = 4.9, std=0.83 year) and 21 typically developing (TD) children (mean age = 5.0, std=0.9 year). Participants engaged in a VR music-based game focused on drumming and xylophone playing, developed by our research team [32]. Within the virtual room, two virtual robots were present: Nima (gray), responsible for drumming, and Sina (red), responsible for the xylophone tasks (Figure 1). The game was divided into three stages, with robot actions controlled through a Wizard of Oz setup:

**Drumming familiarization:** The child was introduced to Nima and performed five drumming imitation tasks using VR controllers without wearing the headset. The robot then instructed the child to put on the headset and proceed to the next stage.

**Xylophone imitation:** Sina performed one of sixteen predefined rhythmic sequences (1–4 beats), which the child had to imitate. If unsuccessful, a simpler task was provided. Sessions were stopped if the child was unable or unwilling to continue.

**Joint attention assessment:** The robot prompted the child to look at wall-mounted boards. The child's gaze and attention level were recorded to evaluate joint attention.

Based on the game design, five exercise categories were included: imitative drumming, single-hit xylophone, multi-hit xylophone, verbal instruction (color-based hitting), and joint attention tasks.

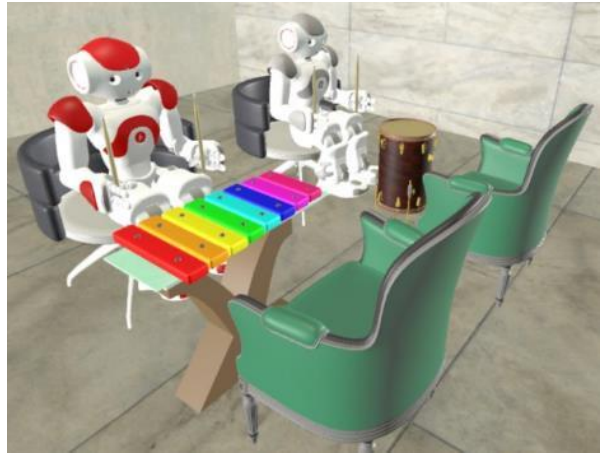

a

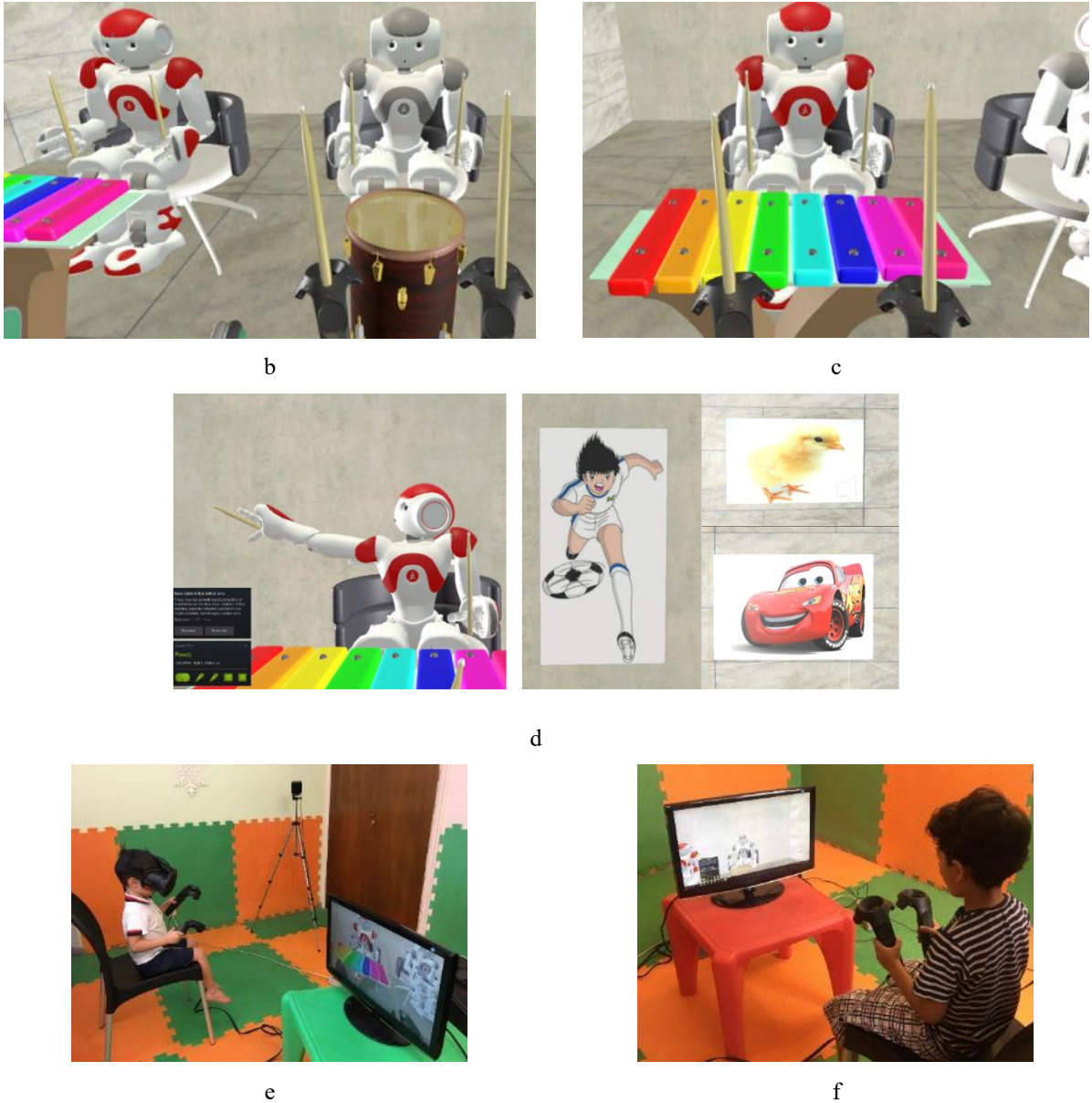

*Figure 1 - Game environment with robots and music tools; a, b, c, and d) Snapshots of the VR room including perspective view of the virtual room, robots, and instruments; Snapshots of the game session: e) a child is playing the xylophone using the headset, and f) a participant is playing the drum without the headset [29].*

## 2. Dataset Construction

During all sessions, two categories of raw data were recorded: 1) video recordings of the gameplay, and 2) positional and angular data of the controllers and headset. The motion signals included 18 parameters per time step, representing three positional and three angular values for each of the headset, right controller, and left controller. Due to incomplete recordings or early termination of sessions, not all data were usable. After excluding invalid cases, data from all TD children and 9 of the 14 ASD participants were retained.

To support analysis, two categories of data were prepared: motion signals and strike events. Motion signals had been recorded in a previous study [32], whereas strike events—comprising the child's drumming or xylophone hits as well as robot cues and strikes—had to be extracted. Direct reconstruction of strikes from raw motion data was unreliable because of variable sampling frequencies. To overcome this, session videos were converted to audio files, and cross-correlation was applied with a similarity threshold of 0.7 and a time tolerance of 0.1 seconds to detect the timing and pitch of notes. The outputs were manually reviewed to correct errors due to sound similarity (e.g., confusion between adjacent notes) or overlap of robot and child strikes. Each validated strike sequence was then categorized according to predefined exercise types.

Segmentation of exercises was based on clear temporal markers. The robot's first instruction or demonstration defined the start of an exercise, and the midpoint between the child's final action and the robot's subsequent cue defined its end. This alignment ensured that strike events and motion signals were synchronized for each exercise.

To standardize temporal resolution, all data were resampled to 16 time-steps per second. Missing rows were interpolated using averaged values, and angular parameters originally stored in the 0–360° range were remapped to –180° to +180°. Exercises were excluded if the child was inattentive, fatigued, repeated prior exercises during a new trial, or if inter-strike intervals were shorter than 1/16 second.

After these preprocessing steps, the dataset for each time step included a timestamp, strike information (by the robot or child), 18 motion parameters, and an exercise label (Table 1). Final counts indicated that although the ASD group size was about half that of the TD group, the number of usable exercises was considerably less. This imbalance might reflect challenges such as reduced compliance, stress, limited attention, slower responses, and quicker fatigue among ASD participants.

*Table 1 - Dataset Overview; the numbers of raw data of the available videos as well as the selected useful one for the current study are mentioned.*

| Exercise Type | TD Children | | ASD Children | |
|---|---|---|---|---|
| | Total | Useful (%) | Total | Useful (%) |
| Drumming | 254 | 196 (76.0%) | 38 | 27 (71.1%) |
| Single-hit Xylophone | 192 | 172 (89.6%) | 40 | 30 (75.0%) |
| Multi-hit Xylophone | 302 | 282 (93.4%) | 46 | 41 (89.1%) |
| Verbal Instruction | 133 | 125 (94.0%) | 36 | 32 (88.9%) |
| Joint Attention | 89 | 72 (80.9%) | 15 | 14 (93.3%) |

## 3. ASD Screening and Classification

The objective of this subsection is to design a neural network–based model for distinguishing between typically developing children and those diagnosed with autism. For this purpose, the classification relies on the activities performed by each child during the VR game environment. Three complementary approaches were considered: 1) classification using only the child's strike events, 2) classification using only motion signals, and 3) classification using both strike events and motion signals in combination. To enable a more rigorous evaluation and ensure reliable model performance, the exercises performed by two randomly selected TD participants (i.e., participants TD-2 and TD-12) and one ASD participant (i.e., participant ASD-2) were completely excluded from the training dataset and reserved solely for the testing phase.

### a) Classification Using Strike Events

This approach used strike events and their timestamps, representing the child's responses to the robot, as input features. The goal was to classify each child into the TD or ASD group (i.e., a binary classification). Since the structure of exercises varied, classification was performed separately for each exercise type.

To achieve this, three classical machine learning models—Random Forest (RF), Support Vector Machine (SVM), and Logistic Regression (LR)—and two deep learning models—a Multi-Layer Perceptron (MLP) and a Long Short-Term Memory network (LSTM)—were implemented. Classical methods were chosen for their interpretability, alignment with previous studies, and suitability for limited datasets, while deep learning methods were applied to capture temporal patterns in the data.

For the classical models, 20% of the data were reserved as the test samples. For the deep learning models, 7-fold cross-validation was adopted, so that about 15% of the data were used for testing in each fold and all samples contributed to training. To address class imbalance, weighted loss functions were applied: the TD class weight was fixed at 1, and the ASD class weight was set to the ratio of TD to ASD samples in each exercise type. Data augmentation methods, such as duplicating ASD samples or adding noise to strike counts, timing, and xylophone bar indices, were also explored, but weighted loss consistently provided the best performance.

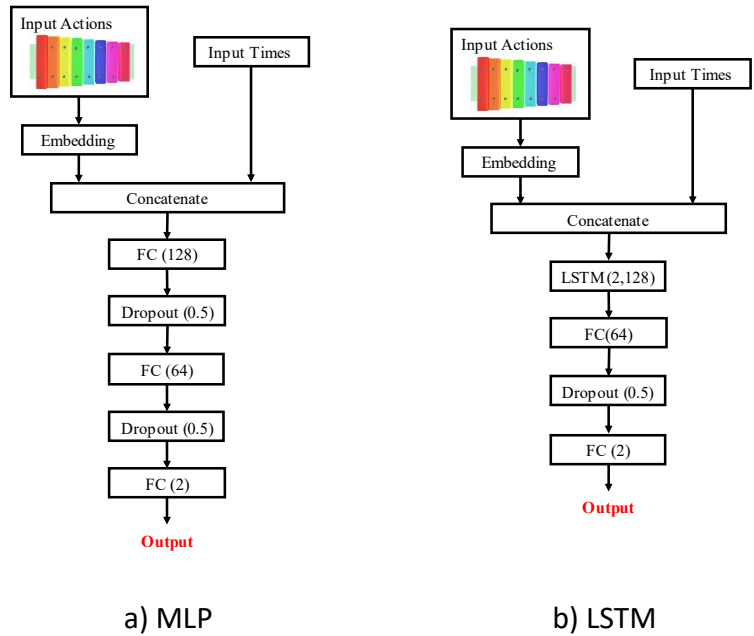

*Figure 2 - Architecture of the deep learning models using strike events.*

As illustrated in Figure 2, for the deep learning models, strike events were embedded and combined with temporal information before being processed by the networks. The MLP and LSTM architectures contained approximately 18,000 and 241,000 trainable parameters, respectively. Both networks used ReLU activation functions, the Adam optimizer ($\beta_1 = 0.9$, $\beta_2 = 0.999$), a learning rate of 0.001, and a decay rate of 0.01. Cross-entropy loss was employed, and training was performed for 100 epochs.

**b) Classification Using Motion Signals**

In this approach, the input features consisted of motion signals derived from the headset and controllers, including 18 positional and angular parameters per time-step. These signals were treated as time series and fed directly into the models. Although the sequences within each individual exercise were of equal length, the overall duration varied between exercises. To ensure consistency, shorter sequences were zero-padded.

Three classical machine learning models—Random Forest (RF), Support Vector Machine (SVM), and Logistic Regression (LR)—were employed, together with three deep learning models: an LSTM network, a CNN+LSTM hybrid, and an encoder-only Transformer. Classical models were included for comparability with existing literature, while the deep models were designed to capture both temporal and spatial dependencies within the motion signals.

All models were trained using 7-fold cross-validation, so that about 15% of the data served as the test samples in each fold while all samples contributed to training. Class imbalance was addressed by applying weighted loss functions, where class weights were proportional to the ratio of TD to ASD samples in each exercise type. The architectures of the deep learning models are illustrated in Figure 3.

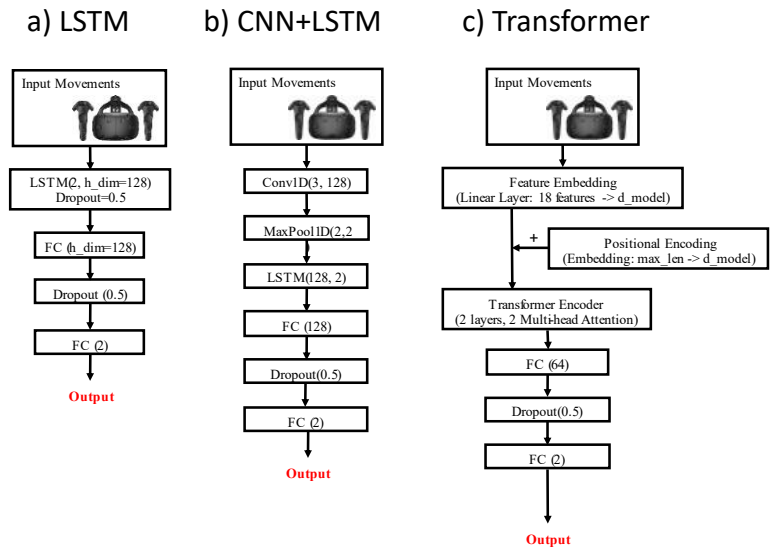

*Figure 3 - Architecture of the deep learning models using motion signals.*

The LSTM, CNN+LSTM, and Transformer architectures contained approximately 224,000, 288,000, and 600,000 trainable parameters, respectively. All networks used ReLU activation functions and were trained with the Adam optimizer ($\beta_1 = 0.9$, $\beta_2 = 0.999$), a learning rate of 0.001, a decay rate of 0.01, and a cross-entropy loss function for 100 epochs. The Transformer model used an embedding dimension of 64, two attention heads per multi-head attention module, two encoder blocks, and positional encoding with a maximum sequence length of 512. The query, key, and value vectors were each set to 32 dimensions per head, producing a combined embedding size of 64.

c) **Classification Using Combined Strike Events and Motion Signals**

In this approach, strike events and motion signals were used jointly to improve classification performance. Timestamps and strike events produced by the child were processed through either an MLP or an LSTM, while the motion signals were simultaneously processed through an LSTM or an encoder-only Transformer. The outputs of these parallel networks were then concatenated and passed into an additional MLP for final classification into TD or ASD groups (i.e., binary classification).

Three combined architectures were designed: MLP-LSTM, MLP-Transformer, and LSTM-Transformer. These networks contained approximately 365,000, 645,000, and 735,000 trainable parameters, respectively. In all cases, ReLU activation functions were used, and training was performed with the Adam optimizer ($\beta_1 = 0.9$, $\beta_2 = 0.999$), a learning rate of 0.001, a decay rate of 0.01, and a cross-entropy loss function. Each network was trained for 50 epochs.

The Transformer component employed an embedding dimension of 64, two attention heads per multi-head attention module, two encoder blocks, and positional encoding with a maximum sequence length of 512. The query, key, and value vectors were set to 32 dimensions per head, yielding a combined embedding size of 64. This architecture allowed the model to integrate temporal strike information with detailed motion dynamics, providing a richer representation for classification.

**4. Behavioral Modeling**

As the main contribution of the paper, the aim of this subsection is to extract behavioral patterns and models from both TD and ASD children during the VR-based music education program. As in the previous subsections, the modeling relied on data collected from children's activities within the VR environment.

For behavioral modeling, a Transformer architecture was employed. The model used a linear embedding of dimension 512 and a positional encoding of 512 dimensions with a maximum length of 5000. The attention mechanism consisted of 8 heads, with each query, key, and value vector set to 64 dimensions. The feed-forward network contained a hidden layer of size 2048 with ReLU activation and a dropout rate of 0.1. The overall architecture comprised 6 encoder blocks and 6 decoder blocks.

Dynamic learning rate scheduling was applied using the Noam strategy. The Adam optimizer was used with an initial learning rate of 0.015 and parameters $\beta_1 = 0.9$, $\beta_2 = 0.98$. The scheduler included a warm-up phase of 400 steps with a multiplier of 1. Mean squared error (MSE) was used as the loss function, and training was conducted for 350 epochs. The model contained approximately 2.65 million parameters.

Training was performed separately for each exercise type. The network input consisted of one-hot encoded robot strikes in the format (number of samples × sequence length × features). In this structure, the number of samples represents the batch size, the sequence length corresponds to the number of time steps, and the number of features indicates the set of unique strikes in that exercise type. For example, in a single-hit xylophone task, if the robot struck the second bar, the input vector would be [0,1,0,0,0,0,0,0], given eight unique bars. In a multi-hit xylophone task with a maximum of three strikes, if the robot performed two strikes (on bars 2 and 7), the input would be [[0,1,0,0,0,0,0,0], [0,0,0,0,0,0,1,0], [0,0,0,0,0,0,0,0]].

The network output followed the structure (number of samples × sequence length × features). Here, the number of samples corresponds to the number of predictions (set to one for a single prediction), the sequence length was defined as the weighted average length of exercises in that category, and the number of features was fixed at 18, representing the positional and angular motion parameters.

After training, the model's predictions were executed within the VR environment to reproduce behaviors under similar conditions. For evaluation, ten graduate students in clinical child psychology, all familiar with ASD, were asked to watch videos of real children and the model-generated behaviors, and then complete a questionnaire to assess the quality of reproduction. For illustration purposes, a representative example from each of the available conditions is presented in this [link](#) (See also Figure 4).

To assess the effectiveness of the proposed behavioral model in simulating realistic child behavior, we conducted a blind questionnaire with domain experts. The central goal was to determine if participants could reliably differentiate between real and synthesized videos, with the hypothesis that a low discrimination rate would indicate a highly realistic model.

For this evaluation, we recruited ten master's degree students in psychology (3 male, 7 female) with an average age of 23.9 years. Participants were first given an overview of the virtual reality music-playing game and children's typical behavior within it. They then watched a total of 20 video clips, unaware of whether the videos were real or synthetically generated. These clips were a balanced music-based scenarios mix of four categories: 1) real typically developing children, 2) real children with ASD, 3) synthesized typically developing children, and 4) synthesized children with ASD.

For each video, participants answered two questions:

- Is the child in the video typically developing or on the autism spectrum?
- Is the observed behavior real or synthesized by the model?

## IV. Results and Discussion

The results of this study are presented in two parts. First, we report the outcomes of the ASD/TD classification experiments based on strike events, motion signals, and their combination. Second, we present the results of behavioral modeling to examine the ability of the proposed model to reproduce children's behaviors under similar conditions.

### 1. Results of Classification

#### a) Classification Using Strike Events

This study employed a range of machine learning and deep learning classifiers to analyze and differentiate between children with and without ASD, based on data derived from physical "hits" during specific exercises (Table 2). It should be noted that for the Joint Attention exercise, the task is purely observational and involves no physical strikes. Therefore, no data related to "hits" was available for this exercise, making it impossible to evaluate classifier performance for this specific activity.

*Table 2 - Results of classification using strike events.*

| Classifier | Metrics | Exercise Type | | | | | Average |
|---|---|---|---|---|---|---|---|
| | | Drumming | Single-hit Xylophone | Multi-hit Xylophone | Verbal Instruction | Joint Attention | |
| Random Forest | Accuracy | 0.75 | 0.80 | 0.91 | 0.85 | --- | 0.81 |
| | F1 score | 0.29 | 0.53 | 0.40 | 0.60 | --- | 0.46 |
| | Sensitivity | 0.30 | 0.56 | 0.33 | 0.50 | --- | 0.43 |
| | Specificity | 0.84 | 0.86 | 0.97 | 0.95 | --- | 0.89 |
| | AUC | 0.59 | 0.78 | 0.79 | 0.78 | --- | 0.74 |
| SVM | Accuracy | 0.77 | 0.83 | 0.81 | 0.81 | --- | **0.81** |
| | F1 score | 0.50 | 0.57 | 0.37 | 0.62 | --- | **0.52** |
| | Sensitivity | 0.56 | 0.44 | 0.60 | 0.67 | --- | **0.57** |
| | Specificity | 0.83 | 0.96 | 0.83 | 0.85 | --- | **0.87** |
| | AUC | 0.68 | 0.69 | 0.83 | 0.79 | --- | **0.75** |
| Logistic Regression | Accuracy | 0.68 | 0.66 | 0.83 | 0.69 | --- | 0.72 |
| | F1 score | 0.36 | 0.25 | 0.40 | 0.33 | --- | 0.34 |
| | Sensitivity | 0.44 | 0.22 | 0.60 | 0.33 | --- | 0.40 |
| | Specificity | 0.74 | 0.81 | 0.85 | 0.80 | --- | 0.80 |
| | AUC | 0.59 | 0.60 | 0.62 | 0.65 | --- | 0.62 |
| MLP | Accuracy | 0.81 | 0.85 | 0.87 | 0.80 | --- | 0.83 |
| | F1 score | 0.23 | 0.59 | 0.67 | 0.63 | --- | 0.53 |
| | Sensitivity | 0.27 | 0.66 | 0.80 | 0.75 | --- | 0.62 |
| | Specificity | 0.84 | 0.89 | 0.89 | 0.83 | --- | 0.86 |
| | AUC | 0.65 | 0.85 | 0.90 | 0.80 | --- | 0.80 |
| LSTM | Accuracy | 0.79 | 0.78 | 0.96 | 0.88 | --- | **0.85** |
| | F1 score | 0.26 | 0.56 | 0.90 | 0.78 | --- | **0.63** |
| | Sensitivity | 0.32 | 0.84 | 1.00 | 0.92 | --- | **0.77** |
| | Specificity | 0.82 | 0.77 | 0.96 | 0.88 | --- | **0.86** |
| | AUC | 0.67 | 0.89 | 1.00 | 0.97 | --- | **0.88** |

Table 2 revealed that in our study, the **Long Short-Term Memory (LSTM)** network was the most effective classifier, achieving the highest average accuracy of **85%**. Its superior performance is further highlighted by its high average AUC (0.88) and an F1 score of 0.63. The LSTM's success can be attributed to its ability to model the temporal dependencies and sequential patterns inherent in the data, which appear to be crucial for distinguishing between the two groups of children.

Among the classical models, the **Support Vector Machine (SVM)** showed the strongest performance, with an average accuracy of **81%**. While this is a respectable result, the SVM's average F1 score of 0.52 and sensitivity of 0.57 suggest that it faced challenges in accurately identifying all positive cases (i.e., children with ASD). In contrast, the Logistic

Regression model consistently underperformed across all metrics, with an average accuracy of only 72% and a notably low F1 score of 0.34.

b)  **Classification Using Motion Signals**

This subsection presents the outcomes of a detailed classification analysis using only motion signals, employing both classic machine learning methods and three deep learning models: LSTM, CNN+LSTM, and a Transformer (Encoder-Only) (Table 3).

*Table 3 - Results of classification using motion signals.*

| Classifier | Metrics | Exercise Type | | | | | Average |
| --- | --- | --- | --- | --- | --- | --- | --- |
| | | Drumming | Single-hit Xylophone | Multi-hit Xylophone | Verbal Instruction | Joint Attention | |
| Random Forest | Accuracy | 0.91 | 1.00 | 0.96 | 1.00 | 0.95 | **0.96** |
| | F1 score | 0.40 | 1.00 | 0.83 | 1.00 | 0.67 | **0.78** |
| | Sensitivity | 0.25 | 1.00 | 0.71 | 1.00 | 0.50 | **0.69** |
| | Specificity | 1.00 | 1.00 | 1.00 | 1.00 | 1.00 | **1.00** |
| | AUC | 0.96 | 1.00 | 0.99 | 1.00 | 0.94 | **0.98** |
| SVM | Accuracy | 0.86 | 0.97 | 0.96 | 0.96 | 0.93 | 0.94 |
| | F1 score | 0.44 | 0.89 | 0.83 | 0.91 | 0.67 | 0.75 |
| | Sensitivity | 0.50 | 1.00 | 0.71 | 0.83 | 0.50 | 0.71 |
| | Specificity | 0.90 | 0.96 | 1.00 | 1.00 | 1.00 | 0.97 |
| | AUC | 0.85 | 0.99 | 1.00 | 1.00 | 1.00 | 0.97 |
| Logistic Regression | Accuracy | 0.83 | 0.97 | 0.98 | 0.96 | 0.93 | 0.93 |
| | F1 score | 0.50 | 0.89 | 0.92 | 0.91 | 0.80 | 0.80 |
| | Sensitivity | 0.75 | 1.00 | 0.86 | 0.83 | 1.00 | 0.89 |
| | Specificity | 0.84 | 0.96 | 1.00 | 1.00 | 0.92 | 0.94 |
| | AUC | 0.85 | 1.00 | 1.00 | 1.00 | 0.96 | 0.96 |
| LSTM | Accuracy | 0.89 | 0.73 | 0.74 | 0.75 | 0.88 | 0.80 |
| | F1 score | 0.00 | 0.46 | 0.48 | 0.57 | 0.48 | 0.40 |
| | Sensitivity | 0.00 | 0.76 | 0.69 | 0.76 | 0.36 | 0.51 |
| | Specificity | 1.00 | 0.74 | 0.79 | 0.76 | 0.97 | 0.85 |
| | AUC | 0.53 | 0.80 | 0.73 | 0.81 | 0.67 | 0.71 |
| CNN + LSTM | Accuracy | 0.89 | 0.70 | 0.68 | 0.74 | 0.76 | 0.75 |
| | F1 score | 0.00 | 0.45 | 0.44 | 0.52 | 0.43 | 0.37 |
| | Sensitivity | 0.00 | 0.79 | 0.74 | 0.64 | 0.57 | 0.55 |
| | Specificity | 1.00 | 0.71 | 0.72 | 0.79 | 0.81 | 0.81 |
| | AUC | 0.48 | 0.71 | 0.72 | 0.77 | 0.73 | 0.68 |
| Transformer | Accuracy | 0.86 | 0.89 | 0.84 | 0.94 | 0.90 | **0.89** |
| | F1 score | 0.57 | 0.68 | 0.58 | 0.86 | 0.64 | **0.67** |
| | Sensitivity | 0.93 | 0.83 | 0.79 | 0.94 | 0.57 | **0.81** |
| | Specificity | 0.81 | 0.90 | 0.83 | 0.94 | 0.96 | **0.89** |
| | AUC | 0.90 | 0.96 | 0.94 | 0.98 | 0.91 | **0.94** |

As it can be seen in Table 3, among the classic machine learning approaches, the **Random Forest (RF)** model demonstrated exceptional performance, achieving the highest average accuracy of **96%**. This is a remarkable result, suggesting that RF can efficiently capture the key features in the motion data for effective classification. While RF had the highest accuracy, it is worth noting a trade-off in its sensitivity (0.69). Other classic methods like SVM and Logistic Regression showed better sensitivity (0.71 and 0.89, respectively), indicating a stronger ability to correctly identify positive cases, even if their overall accuracy was slightly lower (94% and 93%). This nuanced performance highlights the importance of considering multiple metrics beyond just accuracy (Table 3).

In the deep learning category, the **Transformer (Encoder-Only)** model achieved the best results with an impressive average accuracy of **89%** and an average AUC of 0.94. This performance underscores the Transformer's capability to process and understand the complex temporal relationships within the motion data, making it a highly suitable model for this task. It is noteworthy that both the LSTM and CNN+LSTM models underperformed in comparison, with average accuracies of 80% and 75%, respectively. Their particularly low F1 scores (0.40 and 0.37) and sensitivities (0.51 and 0.55) suggest that they struggled to reliably identify positive cases.

Further analysis of the results by exercise type reveals important differences in classifier performance. In the classic methods, the **Drumming** exercise was the most challenging for classification, while the **Xylophone-related** exercises demonstrated the highest separability (It should be noted hitting a drum is much easier than a set of specific bars of a xylophone for children and distinguishing the TD/ASD participants is expected to be more challenging in these kinds of tasks). For the deep learning models, the **Joint Attention** exercise emerged as the best predictor with the highest accuracy, and the **Single-hit Xylophone** exercise showed the highest sensitivity. This indicates that while hit-based data is valuable, certain exercises, like Joint Attention, provide unique motion signals that deep learning models can leverage for superior classification, despite the absence of physical "hits" in the traditional sense.

### c) Classification Using Combined Strike Events and Motion Signals

This subsection explores the performance of the hybrid deep learning models that combine the hit data with the motion signals. This integrated approach aims to leverage the strengths of both data types for more effective classification. Consistent with the previous analysis, the Joint Attention exercise was not evaluated in this section, as it lacks the "hit" data necessary for this combined methodology (Table 4).

The analysis shows that the **MLP-Transformer** model demonstrated the best overall performance in this hybrid approach, achieving an average accuracy of **81%**. This model also secured the highest average F1 score (0.62) and average sensitivity (0.96) among the three hybrid models. The exceptional sensitivity highlights its strong ability to correctly identify positive cases, a crucial factor in the context of screening for autism spectrum disorder. The MLP-Transformer's success suggests that combining a simple feed-forward network (MLP) with the powerful temporal modeling of a Transformer can effectively integrate different data types and learn robust features for this classification task.

*Table 4 - Results of classification using combined strike events and motion signals.*

| Classifier | Metrics | Exercise Type | | | | | Average |
|---|---|---|---|---|---|---|---|
| | | Drumming | Single-hit Xylophone | Multi-hit Xylophone | Verbal Instruction | Joint Attention | |
| MLP - LSTM | Accuracy | 0.76 | 0.73 | 0.76 | 0.75 | --- | 0.75 |
| | F1 score | 0.36 | 0.49 | 0.54 | 0.59 | --- | 0.50 |
| | Sensitivity | 0.52 | 0.85 | 0.85 | 0.87 | --- | 0.77 |
| | Specificity | 0.80 | 0.72 | 0.75 | 0.70 | --- | 0.74 |
| | AUC | 0.71 | 0.89 | 0.88 | 0.88 | --- | 0.84 |
| MLP - Transformer | Accuracy | 0.84 | 0.80 | 0.88 | 0.70 | --- | **0.81** |
| | F1 score | 0.59 | 0.60 | 0.72 | 0.58 | --- | **0.62** |
| | Sensitivity | 1.00 | 0.93 | 0.94 | 0.97 | --- | **0.96** |
| | Specificity | 0.79 | 0.80 | 0.87 | 0.62 | --- | **0.77** |
| | AUC | 0.95 | 0.95 | 0.92 | 0.93 | --- | **0.94** |
| LSTM - Transformer | Accuracy | 0.86 | 0.71 | 0.85 | 0.78 | --- | 0.80 |
| | F1 score | 0.63 | 0.48 | 0.67 | 0.65 | --- | 0.61 |
| | Sensitivity | 1.00 | 0.85 | 0.88 | 0.97 | --- | 0.93 |
| | Specificity | 0.83 | 0.70 | 0.85 | 0.72 | --- | 0.78 |
| | AUC | 0.95 | 0.89 | 0.91 | 0.92 | --- | 0.92 |

By examining the results for each exercise, it is clear that the **Multi-hit Xylophone** exercise yielded the highest average accuracy (83%), while the **Verbal Instruction** exercise had the highest average sensitivity (94%). This shows that each exercise provides unique discriminatory information. The Multi-hit Xylophone exercise's high accuracy suggests it offers data with clearly separable patterns, while the Verbal Instruction exercise's high sensitivity indicates that it generates signals particularly useful for identifying children with ASD.

Overall, the combination of hit and motion data, particularly with the MLP-Transformer model, provides a highly effective method for ASD screening. The results underscore the potential of a multi-modal data approach and highlight the importance of model selection based on both overall accuracy and key metrics like sensitivity, which is critical for clinical applications.

A comparison with existing literature validates our findings. While other studies have reported high accuracies (e.g., Vakadkar et al. at 97.15% [33]), they often rely on large, structured datasets. Our model achieved comparable or better performance on a very small dataset of 30 children using complex, unstructured motion data. Specifically, our model's 96% sensitivity significantly surpasses the results of Farooq et al. (58%) [18], demonstrating a superior capability for detecting the target group. This underscores the robustness and clinical relevance of our proposed methodology.

## 2. Results of Behavioral Modeling

Some sample snapshots of our provided videos (including 3 selected frames of 5 presented real/generated videos) to the psychologists are presented in Figure 4. As we mentioned earlier, for each video, participants answered two questions:

- Is the child in the video typically developing or on the autism spectrum?
- Is the observed behavior real or synthesized by the model?

The mean accuracy for the first question (identifying a typically developing child from one with ASD) was 60%. While this was not the primary focus of our study, it shows that the experts could perform a basic screening with a degree of accuracy, which could potentially be improved with more exposure to the videos.

The most critical finding relates to the second question (distinguishing real from synthesized behavior). The mean accuracy for this question was only **53.5%**. This result is remarkably close to a random guess (i.e., 50%), indicating that the participants, despite their expertise, found it extremely difficult to differentiate between the real and synthesized behaviors. This high level of realism suggests that our model successfully captured the complex and subtle behavioral nuances of both typically developing and ASD children, providing strong evidence for its efficacy in generating realistic, child-like behaviors (Again ,we encourage the respected readers to watch our provided videos in this link).

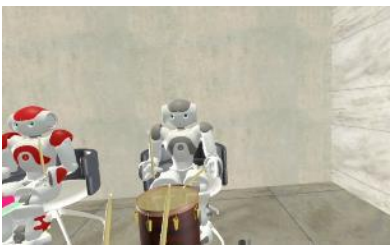
a1
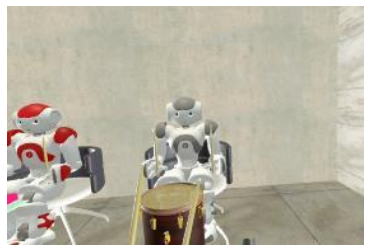
a2
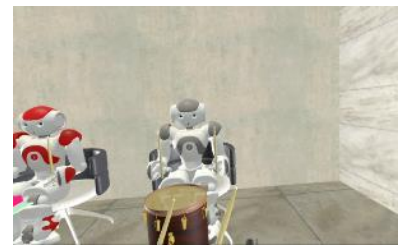
a3
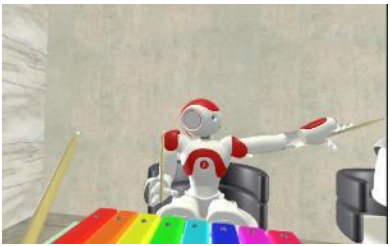
b1
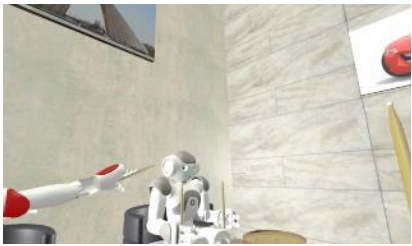
b2
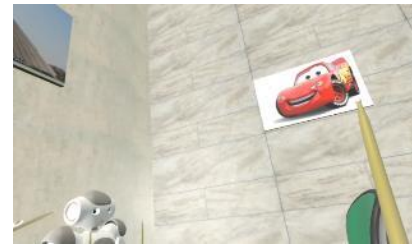
b3

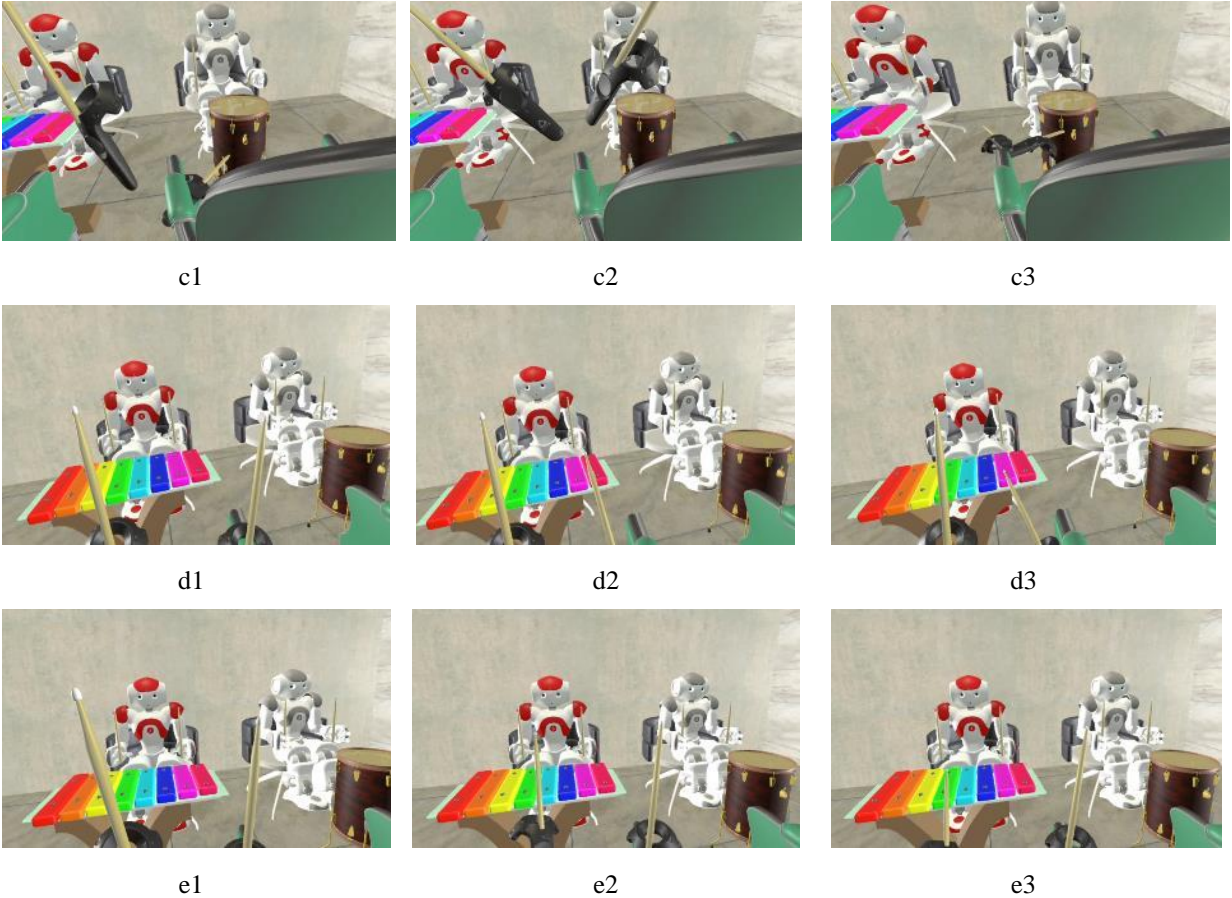

*Figure 4 - Some snapshots (including 3 selected frames) of the real or constructed (predicted) videos of the TD/ASD participants.*

In addition to the mentioned perceptual evaluations, we also examined the stability of the model under variations in the initial conditions (i.e., initial positions and orientations at the first time-step). The model was considered robust if, after altering the initial positions, it consistently generated outputs that remained within acceptable ranges and exhibited human-like motion patterns. Figure 5 illustrates an example of the synthesized motion signals compared with several samples from the existing dataset. As shown, the generated motion remains within the distribution of real signals and demonstrates behavior closely resembling them, further confirming the reliability and stability of our model.

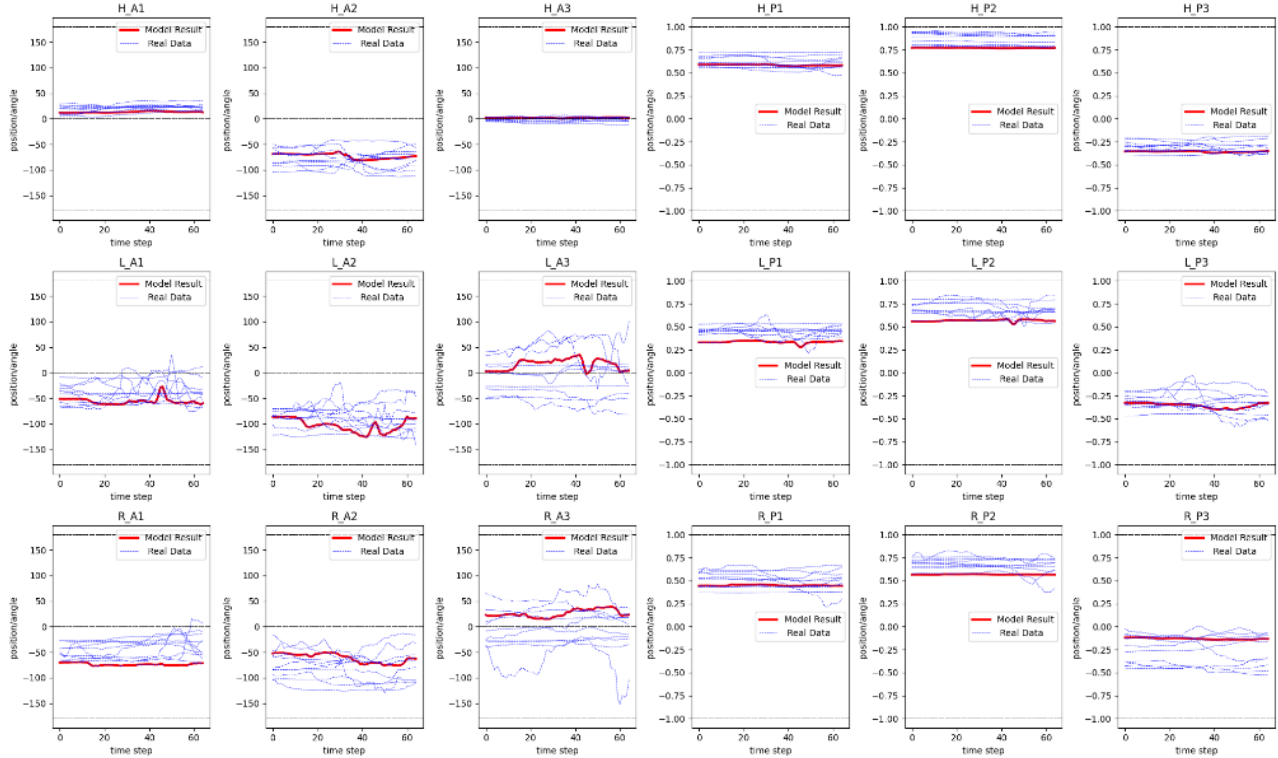

*Figure 5 - Sample of synthesized motion signals compared with several samples from the existing dataset*

### V.     Limitations and Future Work

This study presents several limitations that should be considered when interpreting the findings. The dataset consisted exclusively of children around five years of age, which restricts the generalizability of the results to broader age groups. Moreover, the behavioral exercises administered to the participants were limited and highly specific, potentially narrowing the diversity of observed behaviors and thereby constraining the robustness of the models. Additionally, the dataset lacked essential demographic and clinical variables—including age, gender, and autism severity—which may have influenced the accuracy, sensitivity, and overall effectiveness of the modeling process.

Future research should address these limitations by employing more comprehensive and diverse datasets that include a wider range of age groups and varying degrees of autism severity. Incorporating demographic and individual factors such as age, gender, and clinical severity into the modeling framework could enhance both predictive performance and the precision of personalized interventions. Further advancements may also involve developing smaller, more efficient models suitable for online deployment, eliminating the need for manual "Wizard-of-Oz" operation. Finally, leveraging larger datasets alongside optimized network parameters can enable the design of more powerful and efficient models capable of producing more accurate and impactful outcomes.

## VI. Conclusion

In this study, we evaluated the performance of various models across three distinct data modalities to identify the optimal approach for ASD screening. Our findings highlight a critical trade-off between overall accuracy and sensitivity, the latter being paramount in clinical diagnosis.

For our first approach using only hit data, the **LSTM** model achieved 85% accuracy and 77% sensitivity. The second approach, leveraging only motion signals, saw **Random Forest (RF)** achieve a high 96% accuracy, but with a lower sensitivity of 69%. The **Transformer** model provided a better balance, reaching 89% accuracy and 81% sensitivity. Our third and most effective approach combined both hit and motion data, where the hybrid **MLP-Transformer** model achieved the highest sensitivity of **96%**, along with an accuracy of 81%. We conclude that our **MLP-Transformer** is the most suitable model for this task in our study, as its exceptional sensitivity outweighs the higher accuracy of other models like RF, particularly given the clinical importance of correctly identifying individuals with ASD.

Moreover, our evaluation demonstrates that the proposed behavioral model successfully generates highly realistic and child-like behaviors. The key finding from our expert survey is that participants, despite their professional background, struggled to differentiate between real and synthesized videos, with a discrimination accuracy of only 53.5%. This result, which is remarkably close to random chance, provides strong evidence that the model has successfully captured the complex and subtle nuances of behavior in both typically developing children and those with ASD. While the accuracy for the secondary task of identifying ASD was 60%, the primary goal of realistic synthesis was met with great success, validating the model's potential as a valuable tool for future research and applications in behavioral analysis.

**Acknowledgment**

This work is based upon research funded by the Sharif University of Technology under project No. G4030507. We would like to express our gratitude to all participants and our friends at the Social and Cognitive Robotics Lab. for cooperation and helps with the data collection. We have used ChatGPT for improving the quality of the English writing of the paper.

**Statements & Declarations**

**Conflict of interest**

Author Alireza Taheri has received a research grant from the Sharif University of Technology (Grant No. G4030507)". The authors Armin Tandiseh and Morteza Memari declare that they have no conflict of interest.

**Availability of data and material (data transparency)**

All data from this project (the codes, photos, videos, questionnaire results, etc.) are available in the Social & Cognitive Robotics Laboratory archive.

**Code availability**

All of the codes are available in the archive of the Social & Cognitive Robotics Laboratory.

**Authors' contributions**

All authors contributed to the study's conception and design. Material preparation, data collection, and analysis were performed by Armin Tandiseh. Alireza Taheri supervised this research. The first draft of the manuscript was written by Morteza Memari; and all authors commented on previous versions of the manuscript. All authors read and approved the final manuscript.

**Consent to participate:**

Informed consent was obtained from all individual participants included in the study.